 \documentclass[prd,aps,
preprintnumbers,
twocolumn,
nofootinbib,
floatfix,
superscriptaddress,
longbibliography]{revtex4-2}
\usepackage{graphicx}
\usepackage{epsfig}
\usepackage{rotating}
\usepackage{amssymb}
\usepackage{subfigure}
\usepackage{dsfont}
\usepackage{psfrag}
\usepackage{amsmath,euscript,array,mathrsfs}
\usepackage{tikz-feynman}
\tikzfeynmanset{compat=1.1.0}

\newcommand{\beq}{\begin{equation}}
\newcommand{\eeq}{\end{equation}}
\newcommand{\beqs}{\begin{eqnarray}}
\newcommand{\eeqs}{\end{eqnarray}}

\def\hbar{\hspace{0pt}\raisebox{1pt}{$-$} \hspace{-7pt} h}

\newcommand{\be}{\begin{equation}}
\newcommand{\ee}{\end{equation}}

\newcommand{\bea}{\begin{eqnarray}}
\newcommand{\eea}{\end{eqnarray}}

\def\lbldef#1#2{\expandafter\gdef\csname #1\endcsname {#2}}

\def\href#1#2{#2}

\newcommand{\ber}{\begin{eqnarray}}
\newcommand{\eer}{\end{eqnarray}}

\newcommand{\beqar}{\begin{eqnarray}}

\newcommand{\eeqar}{\end{eqnarray}}

\newcommand{\dsl}
  {\kern.06em\hbox{\raise.15ex\hbox{$/$}\kern-.56em\hbox{$\partial$}}}

\newcommand{\eeqarr}{\end{eqnarray}}
\newcommand{\ZZ}{{\rm \kern 0.275em Z \kern -0.92em Z}\;}

\newcommand{\ignore}[1]{}
\newcommand\SU{\mathrm{SU}}

\newcommand{\cL}{\ensuremath{\mathcal L} }

\newcommand{\cO}{\ensuremath{\mathcal O} }
\newcommand{\Tr}[1]{\ensuremath{\mbox{Tr}\left[ #1 \right]} }

\def\CC{{\mathchoice
{\rm C\mkern-8mu\vrule height1.45ex depth-.05ex
width.05em\mkern9mu\kern-.05em}
{\rm C\mkern-8mu\vrule height1.45ex depth-.05ex
width.05em\mkern9mu\kern-.05em}
{\rm C\mkern-8mu\vrule height1ex depth-.07ex
width.035em\mkern9mu\kern-.035em}
{\rm C\mkern-8mu\vrule height.65ex depth-.1ex
width.025em\mkern8mu\kern-.025em}}}

\def\RR{{\rm I\kern-1.6pt {\rm R}}}

\def\ZZ{{\rm Z}\kern-3.8pt {\rm Z} \kern2pt}
\def\IB{\relax{\rm I\kern-.18em B}}
\def\ID{\relax{\rm I\kern-.18em D}}
\def\II{\relax{\rm I\kern-.18em I}}
\def\IP{\relax{\rm I\kern-.18em P}}

\newcommand{\bear}{\begin{eqnarray}}
\newcommand{\eear}{\end{eqnarray}}

\def\6{\partial}

\def\bea{\begin{eqnarray}}
\def\eea{\end{eqnarray}}

\def\beqx{\begin{displaymath}}
\def\eeqx{\end{displaymath}}

\newcommand{\bmat}{\left(\begin{array}}
\newcommand{\emat}{\end{array}\right)}

\def\cl{{\cal L}}

\def\bo{{\raise-.3ex\hbox{\large$\Box$}}}               % D'Alembertian
\def\face{{\raise.2ex\hbox{$\displaystyle \bigodot$}\mskip-2.2mu \llap {$\ddot
        \smile$}}}                                   % happy face
\def\>{\rangle}                                      %right angle
\def\<{\langle}                                      %left angle

\def\leftrightarrowfill{$\mathsurround=0pt \mathord\leftarrow \mkern-6mu
        \cleaders\hbox{$\mkern-2mu \mathord- \mkern-2mu$}\hfill
        \mkern-6mu \mathord\rightarrow$}        % <--> double differential
\def\dvec#1{\vbox{\ialign{##\crcr
        \leftrightarrowfill\crcr\noalign{\kern-1pt\nointerlineskip}
        $\hfil\displaystyle{#1}\hfil$\crcr}}}           % <--> accent
\def\-{\hphantom{-}}

\newcommand{\pd}[2]{\frac{\partial #1}{\partial #2}} 
\newcommand{\pianndiagOne}[1]{
\begin{tikzpicture}
  [roundnode/.style={circle, draw=black!60, fill=black!6, very thick, inner sep=0.5pt,
  text width=3mm}, %dot/.style={fill=black,circle,minimum size=0.1pt}
  ]
  \begin{feynman}[small]
    \vertex (X) at (0,0);
    \vertex (x1l) at (-1.0,1.0);
    \vertex (x1lp) at (-1.0-.15,1.0-.04){\footnotesize$\pi$};
    \vertex (x2l) at (1.0,1.0);
    \vertex (x2lp) at (1.0+.15,1.0-.05){\footnotesize$\chi$};
    \vertex (x3l) at (1.0,-1.0);
    \vertex (x3lp) at (1.0+.15,-1.0+.05){\footnotesize$\chi$};
    \vertex (x4l) at (-1.0,-1.0);
    \vertex (x4lp) at (-1.0-.15,-1.0+.05){\footnotesize$\pi$};
    \diagram*{
    (x1l) -- [thick, scalar] (X), % -- [thick,  half  left, looseness=1.2, scalar] (Y)  ,
    (x2l) -- [thick] (X),%  -- [thick, quarter left, looseness=1.2, scalar] (Yaux),
    (x3l) -- [thick] (X),  -- [thick, quarter right, looseness=1.2, scalar] (Yaux2)  ,
    (x4l) -- [thick, scalar] (X),  % -- [thick, half right, looseness=1.2, scalar] (Y) ,
    };

    \node[dot] (X1) at (0,0);
  \end{feynman}
\end{tikzpicture}
}

\newcommand{\pianndiagTwo}[1]{
	\begin{tikzpicture}
	[roundnode/.style={circle, draw=black!60, fill=black!6, very thick, inner sep=0.5pt,
		text width=3mm}, %dot/.style={fill=black,circle,minimum size=0.1pt}
	]
	\begin{feynman}[small]
	\vertex (X) at (-0.6,0);
	\vertex (Y) at (0.6,0);
	\vertex (x1l) at (-1.2,1.0);
	\vertex (x1lp) at (-1.2-.15,1.0-.04){\footnotesize$\pi$};
	\vertex (x2l) at (1.2,1.0);
	\vertex (x2lp) at (1.2+.15,1.0-.05){\footnotesize$\chi$};
	\vertex (x3l) at (1.2,-1.0);
	\vertex (x3lp) at (1.2+.15,-1.0+.05){\footnotesize$\chi$};
	\vertex (x4l) at (-1.2,-1.0);
	\vertex (x4lp) at (-1.2-.15,-1.0+.05){\footnotesize$\pi$};
	\diagram*{
		(x1l) -- [thick, scalar] (X), % -- [thick,  half  left, looseness=1.2, scalar] (Y)  ,
		(x2l) -- [thick] (Y),%  -- [thick, quarter left, looseness=1.2, scalar] (Yaux),
		(x3l) -- [thick] (Y),%  -- [thick, quarter right, looseness=1.2, scalar] (Yaux2)  ,
		(x4l) -- [thick, scalar] (X),  % -- [thick, half right, looseness=1.2, scalar] (Y) ,
		(X) -- [thick] (Y),
	};
	
	\node[dot] (X1) at (-0.6,0);
	\node[dot] (Y1) at (0.6,0);
	\end{feynman}
	\end{tikzpicture}
}

\newcommand{\pianndiagThree}[1]{
	\begin{tikzpicture}
	[roundnode/.style={circle, draw=black!60, fill=black!6, very thick, inner sep=0.5pt,
		text width=3mm}, %dot/.style={fill=black,circle,minimum size=0.1pt}
	]
	\begin{feynman}[small]
	\vertex (X) at (0,0.5);
	\vertex (Y) at (0,-0.5);
	\vertex (x1l) at (-1.0,1.0);
	\vertex (x1lp) at (-1.0-.15,1.0-.04){\footnotesize$\pi$};
	\vertex (x2l) at (1.0,1.0);
	\vertex (x2lp) at (1.0+.15,1.0-.05){\footnotesize$\chi$};
	\vertex (x3l) at (1.0,-1.0);
	\vertex (x3lp) at (1.0+.15,-1.0+.05){\footnotesize$\chi$};
	\vertex (x4l) at (-1.0,-1.0);
	\vertex (x4lp) at (-1.0-.15,-1.0+.05){\footnotesize$\pi$};
	\diagram*{
		(x1l) -- [thick, scalar] (X), % -- [thick,  half  left, looseness=1.2, scalar] (Y)  ,
		(x2l) -- [thick] (X),%  -- [thick, quarter left, looseness=1.2, scalar] (Yaux),
		(x3l) -- [thick] (Y),%  -- [thick, quarter right, looseness=1.2, scalar] (Yaux2)  ,
		(x4l) -- [thick, scalar] (Y),  % -- [thick, half right, looseness=1.2, scalar] (Y) ,
		(X) -- [thick, scalar] (Y),
	};
	
	\node[dot] (X1) at (0,0.5);
	\node[dot] (Y1) at (0,-0.5);
	\end{feynman}
	\end{tikzpicture}
}

\newcommand{\pianndiagFour}[1]{
	\begin{tikzpicture}
	[roundnode/.style={circle, draw=black!60, fill=black!6, very thick, inner sep=0.5pt,
		text width=3mm}, %dot/.style={fill=black,circle,minimum size=0.1pt}
	]
	\begin{feynman}[small]
	\vertex (X) at (-0.2,0.6);
	\vertex (Y) at (-0.2,-0.6);
	\vertex (x1l) at (-1.0,1.0);
	\vertex (x1lp) at (-1.0-.15,1.0-.04){\footnotesize$\pi$};
	\vertex (x2l) at (1.0,1.0);
	\vertex (x2lp) at (1.0+.15,1.0-.05){\footnotesize$\chi$};
	\vertex (x3l) at (1.0,-1.0);
	\vertex (x3lp) at (1.0+.15,-1.0+.05){\footnotesize$\chi$};
	\vertex (x4l) at (-1.0,-1.0);
	\vertex (x4lp) at (-1.0-.15,-1.0+.05){\footnotesize$\pi$};
	\diagram*{
		(x1l) -- [thick, scalar] (X), % -- [thick,  half  left, looseness=1.2, scalar] (Y)  ,
		(x2l) -- [thick] (Y),%  -- [thick, quarter left, looseness=1.2, scalar] (Yaux),
		(x3l) -- [thick] (X),%  -- [thick, quarter right, looseness=1.2, scalar] (Yaux2)  ,
		(x4l) -- [thick, scalar] (Y),  % -- [thick, half right, looseness=1.2, scalar] (Y) ,
		(X) -- [thick, scalar] (Y),
	};
	
	\node[dot] (X1) at (-0.2,0.6);
	\node[dot] (Y1) at (-0.2,-0.6);
	\end{feynman}
	\end{tikzpicture}
}

\newcommand{\pianndiagSM}[1]{
	\begin{tikzpicture}
	[roundnode/.style={circle, draw=black!60, fill=black!6, very thick, inner sep=0.5pt,
		text width=3mm}, %dot/.style={fill=black,circle,minimum size=0.1pt}
	]
	\begin{feynman}
	\vertex (X) at (-0.6,0);
	\node[blob, minimum size=0.5cm] (Y) at (0.6,0);
	\vertex (x1l) at (-1.2,1.0);
	\vertex (x1lp) at (-1.2-.15,1.0-.04){\footnotesize$\pi$};
	\vertex (x2l) at (1.2,1.0);
	\vertex (x2lp) at (1.2+.3,1.0-.05){\footnotesize SM};
	\vertex (x3l) at (1.2,-1.0);
	\vertex (x3lp) at (1.2+.3,-1.0+.05){\footnotesize SM};
	\vertex (x4l) at (-1.2,-1.0);
	\vertex (x4lp) at (-1.2-.15,-1.0+.05){\footnotesize$\pi$};
	\vertex (x5l) at (1.9+.05,.7+.05){\footnotesize SM};
	\vertex(x6d) at (1.32-.14,-.6+.12);
	\vertex(x7d) at (1.5-.2,-.2+.04);
	\diagram*{
		(x1l) -- [thick, scalar] (X), % -- [thick,  half  left, looseness=1.2, scalar] (Y)  ,
		(x2l) -- [thick ] (Y),%  -- [thick, quarter left, looseness=1.2, scalar] (Yaux),
		(x3l) -- [thick ] (Y),%  -- [thick, quarter right, looseness=1.2, scalar] (Yaux2)  ,
		(x5l) -- [thick ] (Y),
		(x6d) -- [thick ] (Y),
		(x7d) -- [thick ] (Y),
		(x4l) -- [thick, scalar] (X),  % -- [thick, half right, looseness=1.2, scalar] (Y) ,
		(X) -- [thick, edge label=\(\chi\)] (Y),
	};
	
	\node[dot] (X1) at (-0.6,0);
	\node[dot] (Z1) at (1.32,-.6);
	\node[dot] (Z2) at (1.5,-.2);
	\end{feynman}
	\end{tikzpicture}
}

\preprint{IPPP/24/XX}

\begin{document}
\preprint{IPPP/24/17}	
\title{Dilaton Forbidden Dark Matter}

\author{Thomas Appelquist}
\affiliation{Department of Physics, Sloane Laboratory, Yale University, New Haven, Connecticut 06520, USA}
\author{James Ingoldby}
%\affiliation{Abdus Salam International Centre for Theoretical Physics, Strada Costiera 11, 34151, Trieste, Italy}
\affiliation{Institute for Particle Physics Phenomenology, Durham University, Durham, UK}
\author{Maurizio Piai}
\affiliation{Department of Physics, College of Science, Swansea University, Singleton Park, Swansea, Wales, UK}

%\date{April 11, 2024}
\date{\today}

%\vspace{6mm}

\begin{abstract}
Dilaton effective field theory (dEFT) describes the long distance behavior of certain confining,
near-conformal gauge theories that have been studied via lattice computation. Pseudo-Nambu-
Goldstone bosons (pNGBs), emerging from the breaking of approximate, continuous, internal symmetries, are coupled to an additional scalar particle, the dilaton, arising from the spontaneous breaking of approximate scale invariance. This effective theory has been employed to study possible extensions of the standard model. In this paper, we propose a complementary role for dEFT, as a description of the dark matter of the universe, with the pNGBs identified as the dark-matter particles. We show that this theory provides a natural implementation of the ``forbidden'' dark matter mechanism, and we identify regions of parameter space for which the thermal history of dEFT yields the measured dark matter relic density.
\end{abstract}

\maketitle

%\tableofcontents

\section{Introduction}

The nature of the dark matter of the universe is one of the most important mysteries in fundamental science~\cite{Bertone:2016nfn}. If dark matter takes the form of (non relativistic) new particles, its non-gravitational interactions with the standard model (SM) must be very weak. Yet, current constraints on cold dark matter (CDM) are compatible with a vast range of masses and interaction strengths (see Ref.~\cite{Cooley:2022ufh}, and references therein).

Global fits  of cosmological data within the $\Lambda_{\rm CDM}$ paradigm (the standard cosmological model) determine the present cold dark matter energy density to be $\Omega_{\rm CDM}h^2=0.120 \pm 0.001$~\cite{Planck:2018vyg},\footnote{$h\simeq 0.674(5)$ is Hubble’s constant, $H_0$,  in units of $100\, {\rm km}\, {\rm s}^{-1} {\rm Mpc}^{-1}$, and  $\Omega_{\rm CDM}$ is the CDM energy density normalized to criticality,  $\rho_c\equiv \frac{3 H_0^2}{8\pi G_N}\,=1.878\,\times\,10^{-26} h^2 {\rm kg}\, m^{-3}$~\cite{ParticleDataGroup:2022pth}.} CDM providing about 26-27\% of the energy content of the universe.

A natural mechanism for populating the universe with dark matter  and explaining the present abundance is freezeout; in the early universe, the dark matter was in thermal equilibrium with the SM plasma, decoupling as the universe expanded, and decreasing its interaction rate with the SM. When this mechanism is realized, the relic density, $\Omega_{\rm CDM} h^2$, is determined by the annihilation and replenishment cross sections of dark matter particles.

The dark matter particles could be composite, emerging in a new strongly coupled dark sector. As discussed in Ref.~\cite{Hochberg:2022jfs}, and the review~\cite{Tulin:2017ara}, this possibility may be relevant to anomalies in small-scale structure~\cite{deBlok:2009sp,Boylan-Kolchin:2011qkt}. Furthermore, if the new strong dynamics leads to a first-order phase transition in the early universe, this might yield a relic stochastic gravitational-wave background~\cite{Witten:1984rs,Kamionkowski:1993fg,Allen:1996vm}, detectable in experiments such as LISA~\cite{Caprini:2019egz} or ET~\cite{Maggiore:2019uih}.

One intriguing extension of the freezeout idea is that, within the dark sector, CDM particles annihilate into heavier states, through a process that would be forbidden at zero temperature, but is allowed at finite temperature. This mechanism is known as forbidden dark matter (FDM). See Refs.~\cite{PhysRevD.43.3191,DAgnolo:2015ujb}, and \cite{Delgado:2016umt,DAgnolo:2020mpt} for applications. The thermal suppression in the FDM mechanism allows for realistically large relic densities at freezeout, but with smaller CDM masses and a broad range of self-interaction strengths.

Here, we propose a natural realization of FDM within dEFT. This effective theory extends the conventional chiral Lagrangian with its pNGB fields, $\pi$, to include a dilaton field, $\chi$. It has been extensively studied, in our Refs.~\cite{Appelquist:2017wcg,Appelquist:2017vyy,Appelquist:2019lgk,Appelquist:2020bqj,Appelquist:2022qgl,Appelquist:2022mjb} and in Refs.~\cite{Matsuzaki:2013eva,Golterman:2016lsd,Golterman:2016hlz,Kasai:2016ifi,Hansen:2016fri,Golterman:2016cdd,Fodor:2017nlp,Golterman:2018mfm,Cata:2019edh,Cata:2018wzl,Fodor:2020niv,Golterman:2020utm,Golterman:2021ohm,Fodor:2019vmw,Golterman:2020tdq,LSD:2023uzj,Freeman:2023ket,Zwicky:2023fay,Zwicky:2023krx}. See also the precursors in Refs.~\cite{Migdal:1982jp,Coleman:1985rnk} and Ref.~\cite{Goldberger:2007zk}.  dEFT emerges as the low-energy description of certain underlying gauge theories amenable to lattice studies. One notable example is the $SU(3)$ gauge theory with $N_f = 8$ fundamental Dirac fermions~\cite{Appelquist:2007hu,Deuzeman:2008sc,Fodor:2009wk,Hasenfratz:2014rna,Fodor:2015baa,Aoki:2016wnc,Appelquist:2016viq,Appelquist:2018yqe,Kotov:2021mgp,Hasenfratz:2022zsa,Hasenfratz:2022qan,LatticeStrongDynamicsLSD:2021gmp,LatticeStrongDynamics:2023bqp}. Lattice
studies of this theory have reported evidence for the presence of a light singlet, scalar particle, along with $N^2_f-1$ pNGBs, in the accessible range of fermion masses.

Although the $\SU(3)$ gauge theory provides a UV-complete environment in which to implement the forbidden dark matter dynamics, we restrict attention to only the lightest spin-$0$ states, and describe the dark matter within dEFT, ignoring all other, heavier, composite states. Our approach therefore can apply to other gauge theories that admit a dEFT low-energy
description~\cite{Fodor:2012ty,Fodor:2015vwa,Fodor:2016pls,Fodor:2017nlp,Fodor:2019vmw, Fodor:2020niv}.

We take the dilaton mass, $M_d$, and the pNGB multiplet mass, $M_{\pi}$, to be free parameters, with $M_d>M_{\pi}$. In the $\SU(3)$ gauge theory, throughout the fermion-mass range explored on the lattice, the dilaton {\it is} somewhat heavier than the pNGBs, which are the dark matter particles. This hierarchy is expected to persist down to the chiral limit, since explicit breaking of conformal symmetry remains even there. Finally, we discuss the coupling of the dark sector to the standard model, which must be present to achieve thermal equilibrium between the sectors.

The description of a dark sector in terms of pNGB fields (but no dilaton) has appeared before in the literature on strongly interacting massive particles (SIMPs)~\cite{Hochberg:2014kqa,Hochberg:2015vrg,Hochberg:2018rjs,hochberg2022simp}. There, the depletion takes place through a $3 \rightarrow 2$ process, with the relevant $5$-point interaction
arising from a Wess-Zumino-Witten term. Our dEFT contains $3 \rightarrow 2$ processes even at leading order, but, in the parameter range of interest, they are less important than the dominant $2 \rightarrow 2$ (forbidden) processes. 

Models in which pNGB dark matter is coupled to a dilaton, but which are not of the FDM type, have been identified in the literature~\cite{Kim:2016jbz,Baldes:2021aph,Zhang:2024dgv}.  We also note that an FDM model including a dilaton has appeared recently~\cite{Ferrante:2023bcz}, though in this case the dark matter is not a pNGB. All these approaches are rather different from strongly coupled models in which the dark matter candidate is a baryon~\cite{Appelquist:2015yfa}. 

In Section~\ref{sec:frmwk}, we summarize dEFT for describing the dark matter pNGBs, $\pi$, and the dilaton, $\chi$. The dEFT interactions yield the annihilation process $\pi \pi \rightarrow \chi \chi$, its
inverse, and also the self-interaction among the $\pi$s. In Section~\ref{sec:relden}, we compute the relic density, mapping out the space of allowed masses and coupling strengths. The interaction
with the SM  is discussed in Section~\ref{sec:ints}. We summarize and discuss open questions in Section~\ref{sec:disc}.

\section{Dilaton Effective Field Theory}
\label{sec:frmwk}

Our framework is defined by the three-term Lagrangian
\begin{equation}
	\cl=\cl_\text{SM}+\cl_D+\cl_\text{int}\,,
\end{equation}
where $\cl_\text{SM}$ describes the standard model and $\cL_D$ describes a new composite dark sector with a confinement scale that can be below the electroweak scale. The (highly suppressed) interaction between the two sectors is described by $\cL_\text{int}$. 

The dark-sector Lagrangian, $\cL_D$, is taken to be that of dEFT. Its form, reviewed in Ref.~\cite{Appelquist:2022mjb}, is given by
\begin{multline}
	\cl_D = \frac{1}{2}\partial_{\mu}\chi\partial^{\mu}\chi \, + \frac{F_{\pi}^2}{4}\left(\frac{\chi}{F_d}\right)^2 \, \Tr{\partial_{\mu}\Sigma(\partial^{\mu}\Sigma)^{\dagger}}\\ + \frac{M_{\pi}^2 F_{\pi}^2}{4} \left(\frac{\chi}{F_d}\right)^y \, \Tr{\Sigma + \Sigma^{\dagger}} \, - \, V(\chi) \, ,
	\label{eq:L}
\end{multline}
where $M_{\pi}$ and $F_{\pi}$ are the the mass and decay constant of the composite pNGBs, $\Sigma$ includes the multiplet of pNGB fields with $\Sigma^{\dagger} \Sigma = \mathbf{1}$, and $\chi$ is the dilaton field. The potential, $V(\chi)$, includes a term that explicitly breaks conformal symmetry, and the exponent $y$ is a constant. The dilaton field has a full potential arising from $V(\chi)$ and the third term of Eq.~(\ref{eq:L}):
\begin{align}
W(\chi) \equiv V(\chi) - \frac{M^2_\pi F^2_\pi N_f}{2}\left(\frac{\chi}{F_d}\right)^y\,,
\end{align}
and acquires a nonzero vacuum expectation value (VEV) $\langle\chi\rangle=F_d$ at the potential minimum, breaking approximate scale invariance spontaneously. Expanding about the minimum via the redefinition $\chi  = F_d + \bar{\chi}$ determines the dilaton mass and self-interactions: 
\begin{align}
W(\bar{\chi}) = \text{constant} + \frac{M^2_d}{2}\bar{\chi}^2 + \frac{\gamma}{3!}\frac{M^2_d}{F_d}\bar{\chi}^3+\dots\,,
\end{align}
where $\gamma$ is a constant greater than 2 \cite{Goldberger:2007zk}, which depends on $y$ and the form of the term in $V(\chi)$ that explicitly breaks scale invariance.

To compute the dark-matter relic density, we need to know the cross section for the forbidden process, $\pi\pi\rightarrow\chi\chi$, and its inverse, $\chi\chi\rightarrow\pi\pi$. These depend on the parameters of dEFT: the masses, $M_{d}^2$ and $M_{\pi}^2$, the decay constants, $F_{d}^2$ and $F_{\pi}^2$, the number of pNGBs, which we call $N_{\pi}$, and the dimensionless parameters $y$ and $\gamma$. 

Lattice studies of specific gauge theories, which provide ultraviolet completions for dEFT, yield (model-dependent) information about some of the dEFT parameters. For example, in the $\SU(3)$ gauge theory with $N_f=8$ fundamental fermions, $N_\pi=N^2_f-1=63$ and fits of dEFT to lattice data constrain the parameter $y$ to be close to 2 \cite{Fodor:2019vmw,Golterman:2020tdq,LSD:2023uzj}, a value also supported by theoretical arguments \cite{Zwicky:2023bzk}. For simplicity, we set $y=2$ in the following, although recovering the more general expressions is straightforward. The parameter $\gamma$ is less well determined, and we keep it general. 

The lattice studies of the $N_f=8$ theory explore only a particular range of $M_{\pi}$ values (in lattice units). Throughout this range, $M_d$ is larger than $M_{\pi}$, but of the same order. We take this to be the case in our description of the dark sector. In these data sets, fits to dEFT predictions
indicate that the ratio $F_{\pi}^2/F_{d}^2\simeq 0.1$ \cite{Fodor:2019vmw,Golterman:2020tdq,LSD:2023uzj}, compatible with the expectation that $F_\pi^2/F^2_d\sim N_f^{-1}$.

The effective expansion parameter of dEFT, at momentum scales of order $M_{\pi}$  or smaller, is of order $M_{\pi}^{2} N_{\pi}/ (4 \pi F_{d})^2$. For the range of fermion masses explored in the most recent $N_f = 8$ lattice study~\cite{LatticeStrongDynamics:2023bqp}, $M_{\pi} / F_{\pi} \approx 4$, leading to an expansion parameter of order $0.5$. By reducing fermion masses in the underlying theory, $M_{\pi} / F_{\pi}$ and the effective expansion parameter can be made smaller. We assume $M_{\pi} / F_{\pi}$ is small enough for us to reliably use dEFT to compute the relevant cross sections at freezeout, where the temperatures at play are small compared to $M_{\pi}$. After specifying the parameters of dEFT, there remains only the question of the overall mass scale of the dark sector. We consider a range of possibilities below the electroweak scale.

The dark sector must couple to the standard model, with enough strength to maintain thermal equilibrium between the sectors during freezeout, and yet without enough strength to overwhelm the forbidden mechanism within the dark sector. We discuss this interaction in Section~\ref{sec:ints}, noting that a sizable range of very weak couplings satisfy these constraints.

\section{Relic Density}
\label{sec:relden}

In the FDM mechanism, the population of the  dark matter pNGBs is depleted through a $2\rightarrow2$ scattering process to dilatons. These then decay into SM particles through weak contact interactions to be discussed in Section~\ref{sec:ints}. The relic density is determined by the Boltzmann equation
\begin{align}
	\pd{n_\pi}{t}+3Hn_\pi=&-\langle\sigma_{2\pi\rightarrow2\chi}v\rangle n^2_\pi + \langle\sigma_{2\chi\rightarrow2\pi}v\rangle \left(n_\chi^\text{eq}\right)^2\,,
	\label{eq:boltz}
\end{align}
where $H$ is the Hubble scale and $n_{\pi}$ is the total number density of pNGBs (summed over all flavors). We assume that due to the contact interaction the dilaton remains in thermal equilibrium with the SM during freezeout. For the relevant temperatures, $T<M_\pi,\,M_d$, we represent the pNGB and dilaton number densities using the non-relativistic equilibrium expression
\begin{align}
	n^\text{eq}_{i} = w_i\left(\frac{M_{i}T}{2\pi}\right)^{3/2}e^{-M_{i}/T}\,,
	\label{eq:eq}
\end{align}
where $i=\pi,\,d$, is the pNGB or dilaton, and $w_i$ counts the number of species ($w_d=1$ and $w_\pi=N_\pi$). The quantities $\langle\sigma_{2\pi \rightarrow 2\chi} v\rangle$ and $\langle\sigma_{2\chi \rightarrow 2\pi} v\rangle$ are the thermally averaged cross section and inverse cross section.

The cross section $\langle\sigma_{2\pi \rightarrow 2\chi} v\rangle$, vanishing at $T = 0$ since $M_{\pi} < M_d$, is nonzero at finite $T$, and given in terms of $\langle\sigma_{2\chi \rightarrow 2\pi} v\rangle$ by
\begin{align}
	\langle\sigma_{2\pi\rightarrow2\chi}v\rangle = \frac{(1+\Delta)^3}{N^2_\pi}e^{-2\Delta x}\langle\sigma_{2\chi\rightarrow2\pi}v\rangle\,,
	\label{eq:db}
\end{align}
where $x = M_{\pi}/T$ and $\Delta = (M_d - M_{\pi})/M_{\pi}$. The forbidden rate is exponentially suppressed when $ T \ll M_d - M_{\pi}$. Equation (\ref{eq:db}) can be derived by integrating the two cross sections against the Maxwell-Boltzmann velocity distribution, or by using the principle of detailed balance \cite{DAgnolo:2015ujb}.

Employing Eq.~(\ref{eq:db}) for the thermally averaged $\pi\pi\rightarrow\chi\chi$ cross section in the Boltzmann equation yields
\begin{multline}
	\pd{n_\pi}{t}+3Hn_\pi=-\langle\sigma_{2\chi\rightarrow2\pi}v\rangle\\ \times\left(\frac{n^2_\pi}{N_\pi^2}(1+\Delta)^3e^{-2\Delta x} - \ \left(n_\chi^\text{eq}\right)^2\right)\,.
	\label{eq:boltz2}
\end{multline}

The thermally averaged cross section, $\langle\sigma_{2\chi \rightarrow 2\pi} v \rangle$, can be computed at tree level in dEFT. For the case $y=2$, this cross section is given approximately by
\begin{align}
	\langle\sigma_{2\chi \rightarrow 2\pi} v \rangle = \frac{M^2_\pi N_\pi}{36\pi F^4_d}\sqrt{\Delta(2+\Delta)}(1+\Delta)(5+\gamma)^2\,,
	\label{eq:deftxsec}
\end{align}
where we have neglected corrections of order $1/x=T/M_\pi$. %The four Feynman diagrams that contribute are shown in Fig.~\ref{Fig:2to2}. 
We note that $\gamma$ cannot be too large or else the cross section, Eq.~(\ref{eq:deftxsec}), would exceed unitarity bounds, signaling that the dEFT is outside its domain of validity.

%\begin{figure}[h]
%	 \begin{center}
%	    \pianndiagOne{a}\hspace{10pt}
%		\pianndiagTwo{a}\\\vspace{16pt}
%		\pianndiagThree{a}\hspace{17pt}
%		\pianndiagFour{a}
%         \end{center}
%	\caption{The Feynman diagrams that contribute to the forbidden annihilation of two pNGBs, $\pi$, into two dilatons, $\chi$, within dEFT at leading order.}
%	\label{Fig:2to2}
%\end{figure}

To solve the Boltzmann equation (Eq.~(\ref{eq:boltz2})), we assume a radiation dominated universe so that the Hubble scale, $H$, is given by the expression
\begin{align}
	H = \pi\sqrt{\frac{g(T)}{90}}\frac{T^2}{M_\text{pl}}\,,
	\label{eq:hub}
\end{align}
where $M_\text{pl}=1/\sqrt{8\pi G}$, with $G$ Newton's gravitational constant, and $g(T)$ is a continuous function that counts the effective number of relativistic degrees of freedom.

It is then convenient to recast Eq.~(\ref{eq:boltz2}) in terms of the co-moving number density, $Y_{\pi}=n_\pi/s$, where $s$ is the entropy density given by the expression
\begin{align}
	s(T) = \frac{2 \pi^2}{45} h(T) T^3 \,,
	\label{eq:entropy}
\end{align}
and $h(T)$ is a different function that counts the effective number of degrees of freedom. Free particles with masses $m_i\ll T$ contribute equally to the functions $h(T)$ and $g(T)$. However, these functions differ in general due to mass thresholds and interactions. 

We construct $h(T)$ and $g(T)$ using tabulated values taken from \texttt{micrOMEGAs6.0} \cite{Alguero:2023zol}. Those values are derived from determinations of the functions presented in Ref.~\cite{Drees:2015exa}, themselves obtained using lattice QCD calculations of the equation of state \cite{HotQCD:2014kol}.

In terms of $x=M_\pi/T$, the Boltzmann equation for $Y_{\pi}$ takes the form
\begin{align}
	\frac{dY_\pi}{dx} = -\frac{\xi(M_\pi/x)}{x^2}e^{-2\Delta x}\left(1+\Delta\right)^3\left[Y_\pi^2-\left(Y^\text{eq}_\pi\right)^2\right]\,,
	\label{eq:boltz3}
\end{align}
where $Y_{\pi}^{eq} = n_{\pi}^{eq}(T) /s(T)$, with $n_\pi^\text{eq}(T)$ given by Eq.~(\ref{eq:eq}), while
\begin{align}
	\xi(T) \equiv \frac{2\pi\sqrt{10}M_\pi M_{\text{pl}}}{15N_\pi^2}\langle\sigma v\rangle\frac{h(T)}{\sqrt{g(T)}}\left[1+\frac{1}{3}\frac{d\ln h}{d\ln T}\right]\,.
	\label{eq:xidef}
\end{align}

We solve Eq. (\ref{eq:boltz3}) numerically to obtain the co-moving number density at late times $Y_{\pi}(\infty)$, and from it the relic density of CDM today. A boundary condition must be provided. We take it to be $Y_{\pi}$ at a higher temperature, before freezeout but with $T < M_{\pi}$ ($x > 1$) so that dEFT is applicable and the pNGBs are moving at a nonrelativistic typical speed. At which point we set $Y_{\pi}$ to its non-relativistic, thermal-equilibrium value $Y_{\pi}^{eq}$ (the right-hand side of Eq. (\ref{eq:boltz3}) causes $Y_{\pi}$ to approach $Y_{\pi}^{eq}$ in the range $x > 1$, independently of its behavior at higher temperatures). The requisite late time value $Y_{\pi}(\infty)$ is then insensitive to the behavior of $Y_{\pi}$ at smaller $x$, earlier in cosmological history.

\begin{figure}
	\begin{center}
		\includegraphics[width=0.92\columnwidth]{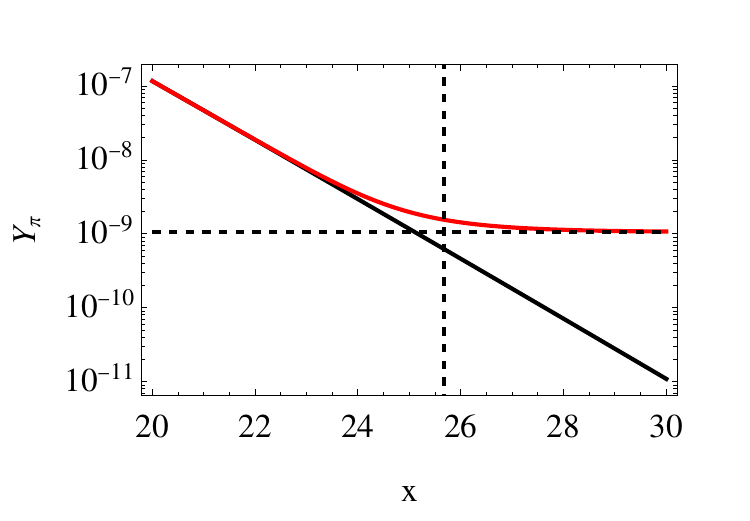}
	\end{center}
\caption{In red, the numerical solution to Eq.~(\ref{eq:boltz3}) for the yield function $Y_\pi$ is shown for the choice $M_\pi=1\,\text{GeV}$ with model parameters $M_\pi/F_\pi=4$, $F^2_\pi/F^2_d=0.1$, $\Delta=0.3$, $\gamma=3$ and $y=2$. The equilibrium yield $Y_{\pi}^\text{eq}$ is plotted as the solid black line for comparison. The freezeout temperature, $x_f$, is represented as the vertical dashed line. The yield at late times (large $x$) is plotted as the dashed horizontal line.}
\label{Fig:yield}
\end{figure}

We plot $Y_{\pi}(x)$ in Fig.~\ref{Fig:yield}, for an illustrative choice of dEFT parameters. Its qualitative behavior is similar for a wide range of parameter choices. The late-time value, $Y_{\pi}(\infty)$, which depends on dEFT parameters, determines the relic density of CDM today through the expression
\begin{align}
	\Omega_{\rm CDM}h^2=\frac{M_\pi s_0 Y_\pi(\infty)}{\rho_c/h^2}\,,
	\label{eq:relY}
\end{align}
where $s_0=2970\text{ cm}^{-3}$ is the entropy density today. Setting the relic density to its observed value $\Omega_{CDM}h^{2} = 0.120 \pm 0.001$ \cite{Planck:2018vyg} allows us to derive a constraint on the allowed parameter space of dEFT.

%The form of this constraint depends on the parameter sensitivity of $Y_{\pi}(\infty)$, determined through the numerical solution to Eq. (\ref{eq:boltz3}). In Fig.~\ref{Fig:mrange},  we plot the allowed values of $M_{\pi}$ versus $\Delta$ for three fixed values of the ratio $M_{\pi}/F_{\pi}$. These correspond to three fixed values of the effective expansion parameter of dEFT.

\begin{figure}
	\begin{center}
		\includegraphics[width=0.8\columnwidth]{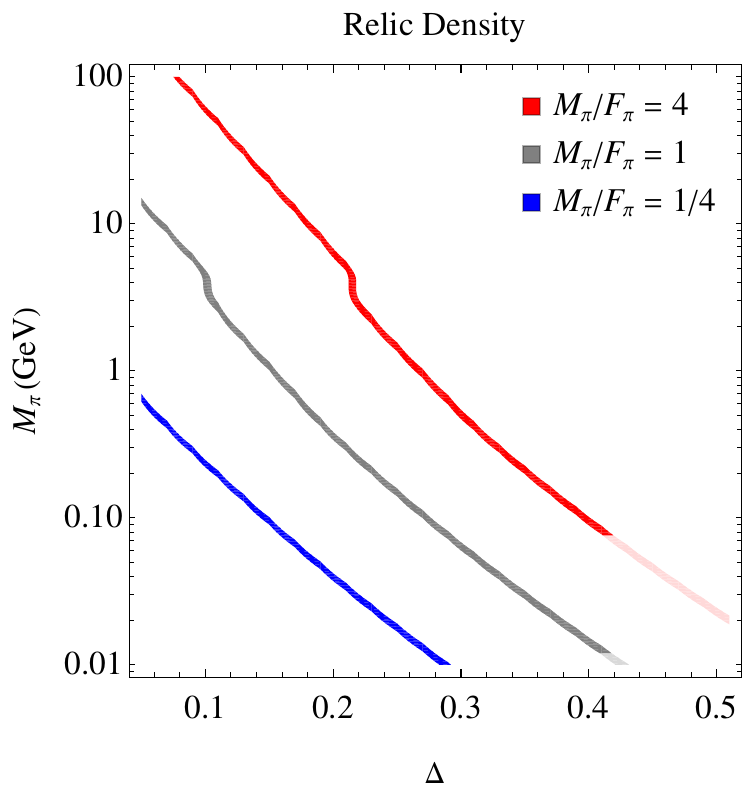}
	\end{center}
	\caption{The bands indicate the parts of the parameter space for which the dark-matter relic density is within 10\% of its observed value. The red, gray and blue colors correspond to three values of the quantity $M_\pi/F_\pi$, which determines the dark matter self interaction coupling. In the bottom right corner, there are portions of the bands shaded in paler colors, for which the dark-matter mass falls below the lower bound shown in Eq.~(\ref{eq:aprxlowerbnd}), setting $\sigma_\text{max}/M_\pi=2\,\text{cm}^2/g$.  For reference, we have also set $F^2_\pi/F^2_d=0.1$, $\gamma=3$ and $y=2$.}
	\label{Fig:mrange}
\end{figure}

The form of this constraint depends on the parameter sensitivity of $Y_{\pi}(\infty)$, determined through the numerical solution to Eq.~(\ref{eq:boltz3}). A range of possible values of the dark-matter mass scale $M_{\pi}$ emerges depending on the dEFT parameters. In Fig.~\ref{Fig:mrange}, we plot the value of $M_{\pi}$ versus $\Delta$ for three values of the ratio $M_{\pi}/F_{\pi}$, corresponding to three values of the effective expansion parameter. For a wide range of dEFT parameters, we are led to a mass scale $M_\pi$ in the broad GeV range. 

Figure~\ref{Fig:mrange} can be understood qualitatively by inspection of Eqs.~(\ref{eq:boltz3}) and (\ref{eq:xidef}), which determine $Y_{\pi}(\infty)$.  It can be argued that this quantity grows approximately linearly with $M_{\pi}$ and exponentially with $\Delta$. This behavior can be exhibited explicitly if Eq.~(\ref{eq:boltz3}) is solved analytically, which is possible if the evolution takes place at temperatures $T$ where $g(T)$ and $h(T)$ are approximately constant. These functions however vary rapidly for temperatures around the QCD confinement scale, giving rise to the kinks visible in Fig.~\ref{Fig:mrange} for $M_\pi\sim 3$ GeV.

As indicated by the shading in Figure~\ref{Fig:mrange}, the range of $M_{\pi}$ values is further constrained, from below, by the strength of elastic dark matter scattering. The behavior of merging galaxy clusters limits the size of this cross section to lie below the bound \cite{Robertson:2016xjh,Wittman:2017gxn}
\begin{align}
	\frac{\sigma}{M_\pi}\le \frac{\sigma_\text{max}}{M_\pi}\approx2\,\text{cm}^2/\text{g}\,.
	\label{eq:sigbound}
\end{align}
The elastic cross section has the form
\begin{multline}
	\sigma=\frac{M^2_\pi}{128\pi F^4_\pi N^2_\pi}\bigg[\frac{a^2}{4}-\frac{b^2M^2_\pi F^2_\pi}{(4M^2_\pi-M^2_d)F^2_d}\\+\frac{c^2M^4_\pi F^4_\pi}{(4M^2_\pi-M^2_d)^2F^4_d}\bigg]\,,
	\label{eq:pipiscat}
\end{multline}
where the parameters $a^2$, $b^2$, and $c^2$ are listed in Table~\ref{Tab:coeffs}. In the regime $F_{\pi}^2 \ll F_{d}^2$, the contribution of the dilaton to Eq.~(\ref{eq:pipiscat}) is negligible and the terms proportional to $b^2$ and $c^2$ can be dropped. Our thermal cross section then coincides with the result of Ref.~\cite{Hochberg:2014kqa}. For the $N_f=8$ theory, the lower bound on $M_{\pi}$ becomes
\begin{align}
	M_\pi\ge11.8\text{ MeV }\left(\frac{M_\pi}{F_\pi}\right)^{4/3}\left(\frac{2\,\text{cm}^2/\text{g}}{\sigma_\text{max}/M_\pi}\right)^{1/3}\,.
	\label{eq:aprxlowerbnd}
\end{align}

\begin{table}[b]
	\centering
	\renewcommand\arraystretch{1.4}
	\begin{tabular}{| c || c |}
		\hline
		$a^2 N^2_f$ & $8(N^2_f-1)(3N^4_f-2N^2_f+6)$ \\
		$b^2 N_f$ & $64(N^2_f-1)(2N^2_f-1)$ \\
		$c^2$ & $256(N^2_f-1)^2$\\
		\hline
	\end{tabular}
\caption{Coefficients that appear in the $2\pi\rightarrow2\pi$ scattering cross section in Eq.~(\ref{eq:pipiscat}).}
\label{Tab:coeffs}
\end{table}

Using the numerical solution of Eq.~(\ref{eq:boltz3}), we can also determine the freezeout temperature, $T_f$. We take this to be the temperature below which the comoving number density of pNGBs, $Y_\pi(x)$, begins to rise significantly above the value it would have if it were in thermal equilibrium with the rest of the SM $Y_\pi^\text{eq}(x)$, indicating that the pNGBs have decoupled from the SM bath. Following the convention of \cite{belanger2013micromegas}, we take $Y_\pi(x_f)=2.5Y^\text{eq}_\pi(x_f)$, where $x_f=M_\pi/T_f$. We plot $x_f$ as a vertical dashed line in Fig.~\ref{Fig:yield}.

\begin{figure}
	\begin{center}
		\includegraphics[width=0.9\columnwidth]{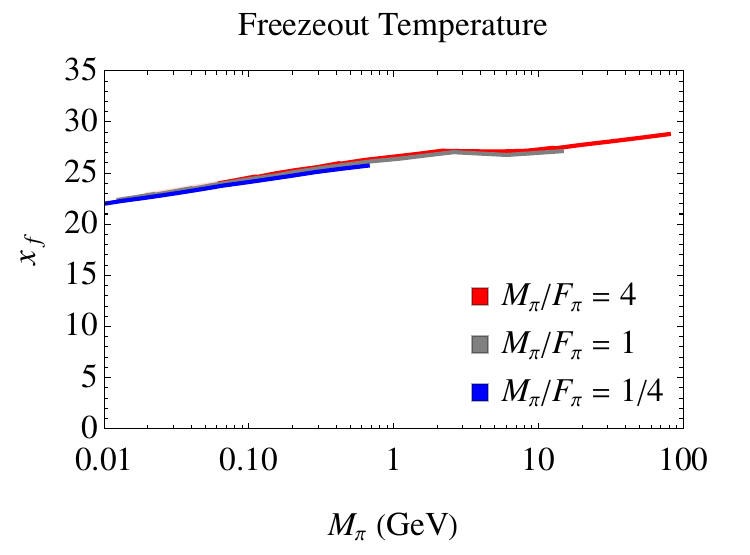}
	\end{center}
	\caption{Plot of $x_f=M_\pi/T_f$ as a function of the dark-matter mass, $M_\pi$, for different choices of $M_\pi/F_\pi$. The value of $\Delta$ has been adjusted to ensure that the dark matter relic density is equal to its observed value. For reference, we have also set  $F^2_\pi/F^2_d=0.1$, $\gamma=3$ and $y=2$.}
	\label{Fig:trange}
\end{figure}

In Fig.~\ref{Fig:trange}, we show the dependence of $x_f$ on $M_\pi$. The figure indicates this dependence is approximately linear when viewed using a log scale for $M_\pi$, which can be explained using Eq.~(\ref{eq:relY}). Since $\Omega_\text{CDM}h^2$ is being fixed to its observed value, it implies that $M_\pi Y_\pi(\infty)=0.4\,\text{eV}$. Furthermore, $Y_\pi(\infty)\sim n^\text{eq}_\pi(T_f)/s(T_f)\sim x_f^{3/2}e^{-x_f}$ up to constants. Inverting this relationship approximately yields $x_f \sim \text{const} + \log(M_\pi/\text{GeV})$. Note that $n^\text{eq}_\pi(T)$ carries no direct dependence on the ratio $M_\pi/F_\pi$, explaining the observed lack of variation with this quantity. The kinks visible around $M_\pi\approx3$ GeV arise due to the abrupt change in relativistic degrees of freedom $h(T)$ around the QCD confinement scale.

\subsection{Subdominant $3\rightarrow 2$ Processes}

In addition to the $2\rightarrow2$ annihilations of dark-matter pNGBs to dilatons, the population of pNGBs can also be depleted through $3\rightarrow 2$ processes, as in the composite SIMP models of Refs.~\cite{Hochberg:2014kqa,Hochberg:2015vrg,Hochberg:2018rjs,hochberg2022simp}. There, with the dark matter described only by pNGBs (no dilaton), the relevant five-point interaction arises from the Wess-Zumino-Witten term. In our framework, we must also account for the $3\rightarrow2$ process, $\pi\pi\pi \rightarrow \chi\pi$, with an outgoing dilaton, which arises at leading order in dEFT. It is unsuppressed thermally, as long as $M_d < 2 M_{\pi}$.

The relative importance of this process during freezeout is given by the ratio
\begin{align}
	R = \left.\frac{n^\text{eq}_\pi\langle\sigma_{3\pi\rightarrow\chi\pi} v^2\rangle}{\langle\sigma_{2\pi\rightarrow2\chi} v\rangle}\right|_{x=x_f}\,.
\end{align}
The denominator $2 \rightarrow 2$ cross section is given by Eqs.~(\ref{eq:db}) and (\ref{eq:deftxsec}), which at freezeout, with $\Delta$ and $\gamma$ in the relevant range, is of order
\begin{align} 
	\langle\sigma_{2\pi\rightarrow2\chi} v\rangle  \approx \frac{M^2_\pi e^{-2\Delta x_f}}{F^4_d N_\pi}\,.
\end{align}
The numerator $3 \rightarrow 2$ cross section, with mass dimension $-5$, arises from a tree level amplitude of order  $M^2_\pi/(F_\pi^{2} F_{d})$. The phase-space integral is similar to that of the $2 \rightarrow 2$ cross section. Thus we expect the cross section to be roughly of order
\begin{align}
		\langle\sigma_{3\pi\rightarrow\chi\pi} v^2\rangle  \approx \frac{M_\pi}{F^4_\pi F^2_d N_\pi}\,.
\end{align}
Using Eq.~(\ref{eq:eq}) for the equilibrium number density, we have
\begin{align}
	R\approx N_\pi \frac{M^2_\pi F^2_d}{F^4_\pi} x_f^{-3/2} e^{-(1-2\Delta)x_f}\,.
\end{align}
Thus $R$ is exponentially suppressed due to the factor $n_{\pi}^{eq}$ associated with the $3\rightarrow2$ process, provided that $\Delta < 0.5$. This restriction, implemented in Fig.~\ref{Fig:mrange}, is satisfied by the LSD lattice data, and for a range of $M_{\pi}$
values closer to the chiral limit. We note that the relative importance of $2\rightarrow 2$ and $3\rightarrow 2$ processes has also been discussed in a different set of models \cite{Cline:2017tka,Bernreuther:2019pfb,Bernreuther:2023kcg,Chu:2024rrv,Pomper:2024otb}.

\section{Interaction with the SM}
\label{sec:ints}

The FDM mechanism and the form of Eq.~(\ref{eq:boltz}) incorporate thermal equilibrium between the standard model and the dark-sector dilaton. To provide the necessary coupling between the dilaton and the standard model, we introduce weak contact interactions of the form
\begin{align}
	\cL_\text{int} = \epsilon F_d^{4-d_\text{SM}}\left(\frac{\bar{\chi}}{F_d}\right)\cO_\text{SM} \,,
	\label{eq:contact}
\end{align}
where $\bar{\chi} \equiv \chi - F_d$, and $\cO_{SM}$ is an SM-singlet operator with engineering dimension $d_\text{SM}$. We envision that this interaction arises at high scales, well above those of the dEFT and the SM, leading to a small value for the dimensionless parameter $\epsilon$. We do not describe the operators $\cO_{SM}$ in detail here, noting only two key features. Firstly, they are built from SM fields that are sufficiently light after electroweak symmetry breaking to contribute to the decay of the dilaton. Secondly, since the dilaton, $\chi$, emerges within only the dark sector, they are not directly associated with the possibility of spontaneous breaking of approximate scale symmetry in the SM. Thus the operators $\cO_{SM}$ are not constrained to be related to the trace of the energy momentum tensor of the SM.

We denote the inclusive decay rate of the dilaton into SM particles as $\Gamma_{\chi \rightarrow \text{SM}}$. It must be large enough to ensure that the dark sector and the SM reach thermal equilibrium well before the freezeout process begins. We take this to be at a common temperature of order $M_{\pi}$, just as the dark matter is becoming non-relativistic. The lower limit is taken to be
\begin{equation}
	\Gamma_{\chi\rightarrow\text{SM}} \gtrsim H_{T=M_\pi}\,.
	\label{eq:lb}
\end{equation}

It is also important that the direct annihilation of pNGBs to SM particles through a virtual dilaton does not dominate the forbidden annihilations. This leads to the inequality
\begin{equation}
	\langle\sigma_{2\pi \rightarrow \text{SM}}v\rangle \lesssim \langle\sigma_{2\pi \rightarrow 2\chi}v\rangle\,,
	\label{eq:xsecbound}
\end{equation}
where the strongest bound comes when the right hand side is taken to be the cross section at roughly the freezeout temperature. There, $\langle\sigma_{2 \pi \rightarrow 2\chi}v\rangle \approx H_{T = T_f} /n_{\pi}^\text{eq} (T_f)$.

Equation~(\ref{eq:xsecbound}) leads to an upper bound on $\Gamma_{\chi \rightarrow SM}$ since the dominant contribution to the left-hand side comes from an s-channel dilaton exchange%as depicted in Fig.~\ref{Fig:smann}
. The general ``resonance'' expression for this cross section is 
\begin{align}
	\langle\sigma_{2\pi \rightarrow \text{SM}}v\rangle = \frac{M^5_d\Gamma_{\chi\rightarrow\text{SM}}}{4N_\pi M^2_\pi F^2_d\left(4M^2_\pi-M^2_d\right)^2}\,,
	\label{eq:res}
\end{align}
where we have assumed a narrow width and taken $y = 2$, as before. Since $2M_{\pi}>M_d$ throughout the parameter space we consider, the dilaton is unable to decay to two pNGBs, and is well off shell. However, the resonance formula may still be applied in this case \cite{Ibe:2008ye}.

We combine Eqs.~(\ref{eq:lb}) - (\ref{eq:res}), employing Eq.~(\ref{eq:db}) for $\langle\sigma_{2\pi \rightarrow 2\chi} v\rangle$. This leads to the order-of-magnitude double-sided bound
\begin{align}
	H_{T=M_\pi} \lesssim \Gamma_{\chi\rightarrow\text{SM}} \lesssim H_{T=T_f}\frac{M_\pi N_\pi F^2_d}{n_\pi^\text{eq}(T_f)}\,,
\end{align}
where we have disregarded numerical factors in the upper bound, anticipating that its relative largeness is determined dominantly by the exponential suppression of $n_{\pi}^\text{eq}$ in the denominator. The two bounds insure a sizable allowed range for the decay rate $\Gamma_{\chi \rightarrow SM}$, but restricted to very small values relative to $M_{d} \approx M_{\pi}$. 

If the decay rate  $\Gamma_{\chi \rightarrow SM}$ arises from a single interaction of the form Eq.~(\ref{eq:contact}), we can roughly bound the parameter $\epsilon$. For the case $d_\text{SM} = 3$, and taking the decay to be into two light SM states, we have $\Gamma_{\chi \rightarrow \text{SM}}\approx\epsilon^2 M_d/ 16 \pi$. For the choice $M_{\pi} / F_{\pi} = 1$ for example, and taking  $M_{\pi}\approx M_{d}=1$ GeV, the forbidden mechanism is viable providing $10^{-9} \lesssim \epsilon \lesssim 10^{-4}$. Both the lower and upper bounds grow with $M_{\pi}$, but remain very small for the range of allowed mass values shown in Fig.~\ref{Fig:mrange}.

The magnitude of the weak dilaton-SM interactions in Eq.~(\ref{eq:contact}), the structure of the operators $\cO_\text{SM}$, and the mass scale $M_{\pi}$ will determine possible experimental signatures of the dark matter. These could include dilaton production and decay in collider searches as well as signals in direct and indirect detection experiments. It will be valuable to develop predictions for these phenomena, which will involve further modeling of the dilaton-SM interactions.

\section{Summary and Discussion}
\label{sec:disc}

We have proposed a description of composite dark matter based on dilaton effective field theory (dEFT). The dark-matter particles are pseudo-Nambu-Goldstone bosons (pNGBs) arising from an underlying, near conformal gauge theory. Lattice studies of near conformal gauge theories have led also to a relatively light scalar particle, identified as an approximate dilaton in dEFT. In the fermion-mass range of the lattice studies, the dilaton mass $M_d$ is of order the pNGB mass $M_{\pi}$, but somewhat larger.

Thus the dEFT provides a natural implementation of the forbidden dark-matter framework~\cite{PhysRevD.43.3191,DAgnolo:2015ujb}, in which freezeout of the dark-matter population is described by $2 \rightarrow 2$ scattering to somewhat heavier particles. The process takes two pNGBs to two dilatons (which then transition rapidly to SM particles).  Forbidden at zero temperature by the dilaton-pNGB mass gap, this process becomes allowed but suppressed at finite temperatures small compared to $M_{\pi}$.

The dark-sector dEFT we have employed derives from an underlying $\SU(3)$ gauge theory with $N_f = 8$ Dirac fermions. After describing the dEFT, along with an effective coupling to the SM, we have discussed the computation of the relic dark-matter density in terms of the late-time, co-moving number density $Y_{\pi}(\infty)$ (Eq.~(\ref{eq:relY})). Setting the relic density to its observed value has then allowed us to derive a constraint on the allowed parameter space of the dEFT.

We have adopted the value $F_{\pi}^2 / F_d^2 \sim 0.1$ for the ratio of pNGB and dilaton decay constants, as suggested by the lattice computations. We have then considered several values for the ratio $M_{\pi} / F_{\pi}$, which determines the interaction strength of the dEFT. One, $M_{\pi} / F_{\pi} = 4$, is typical of the current range of lattice data. It leads to an effective dEFT coupling strength of order $0.5$ (at the edge of weak coupling). The other, smaller values for $M_{\pi} / F_{\pi}$ will emerge from lattice studies closer to the chiral limit, and place the dEFT further inside the weak coupling range.

With either of these choices for $M_{\pi}/F_{\pi}$, a correlated set of values for $M_{\pi}$ and $\Delta \equiv (M_d - M_{\pi})/M_{\pi}$ yields the measured relic dark-matter density, as shown in Fig.~\ref{Fig:mrange}. For the case $M_{\pi} / F_{\pi} = 1$, for example, $M_{\pi}$ can range up to roughly $10$ GeV without the need to fine tune $\Delta$ close to $0$. It can range down to roughly $10$ MeV, keeping $\Delta$ below $0.4$, ensuring that the forbidden process dominates the SIMP ($3 \rightarrow 2$) process. Freezeout of the dark matter takes place at temperatures of order $M_{\pi}/25$, where the dEFT is reliable and the dark matter is non-relativistic. 

We have briefly discussed the effective interaction that must be present between the dEFT dark sector and the particles of the SM, observing that a range of very weak couplings are allowed, indicating that the interactions arise from new physics at a very high scale. The detailed form of these interactions will dictate possible experimental signatures of the dark matter.

The particular dEFT we have employed to describe the dark sector, linked reassuringly to a specific underlying gauge theory studied on the lattice, should be regarded as only one possibility. Other underlying gauge theories that yield a light scalar could lead to other dEFTs describing all the emergent light bound states.

\vspace{1.0cm}
\begin{acknowledgments}
	
	We gratefully acknowledge Suchita Kulkarni for providing comments on earlier drafts of our manuscript. \\
	
	The work of MP has been supported in part by the STFC Consolidated Grants No. ST/P00055X/1, ST/T000813/1, and ST/X000648/1. MP received funding from the European Research Council (ERC) under the European Union’s Horizon 2020 research and innovation program under Grant Agreement No. 813942. \\
	
	\noindent\textbf{Open Access Statement} -- For the purpose of open access, the authors have applied a Creative Commons Attribution (CC BY) licence to any Author Accepted Manuscript version arising.
	
	\noindent\textbf{Research Data Access Statement} -- No new data were generated for this manuscript.

\vspace{1.0cm}

\end{acknowledgments}
%\newpage
%%%%%%%%%%%%%%%%%%%%%%%%%%%%%%%%%%%%%%%%
%%%%%%%%%%%%%%%%%%%%%%%%%%%%%%%%%%%%%%%%

\bibliography{DilatonDM}

\end{document}